\documentclass{nature_fig}

\usepackage{graphicx}
\usepackage{color}
\usepackage[normalem]{ulem}
\usepackage{epstopdf}
\usepackage{amsmath}
\usepackage{amssymb}
\usepackage{color}
\usepackage{caption}

\title{Nanomechanical system meets ultra-small, robust, and ultra-low-powered digital communication receiver}

\author{Keita Funayama$^{1,2}$, Hiroya Tanaka$^1$, Jun Hirotani$^2$, Keiichi Shimaoka$^1$,  Yutaka Ohno$^{2,3}$, \& Yukihiro Tadokoro$^1$}

\begin{document}

\maketitle

\begin{affiliations}
 \item TOYOTA Central R\&D Labs., Inc., Nagakute, Aichi 480-1192, JAPAN.
 \item Department of Electronics, Nagoya University, Nagoya, Aichi 464-8601, JAPAN.
 \item Institute of Materials and Systems for Sustainability, Nagoya University, Nagoya, Aichi 464-8601, JAPAN. 
\end{affiliations}

\begin{abstract}
Nanomechanical systems offer a versatile platform for both fundamental science and industrial applications\cite{Roukes2001, Bunch2007, Sazonova2004}.
Resonating vibration has been demonstrated to enable an ultra-sensitive detection of various physical quantities\cite{Jensen2008, Lee2010, Moser2013, Moser2014, Puller2013}, with emerging applications including signal processing\cite{Jensen2007, Hatanaka2017, Mahboob2008, Torrejon2017}, biological detection\cite{Eom2011, Kosaka2014} and fundamental tests of quantum mechanics\cite{Chu2017, Hong2012, Connell2010, Singh2014, Tavernarakis2018}.
It has also been shown that the mechanical vibration of a nanoscale cantilever can be used to detect electromagnetic analogue-modulated waves\cite{Jensen2007}.
This concept is quite exciting; it opens the door to the possibility of fabricating an ultra-small transceiver by realising extreme scale-down of receiving antenna\cite{Atakan2010}.
Indeed, aiming for digital data transfer with such small wireless terminals, some research groups have attempted to detect digitally modulated signals with nanomechanical systems\cite{tadokoro2018, tadokoro2018scirep, Gouttenoire2010}.
However, signals obtained with nanoscale receivers are so weak that the data transfer often fails\cite{tadokoro2018}; to the best of our knowledge, no successful practical demonstration has yet been reported.
Here, we present the first experimental demonstration of the use of nanomechanical systems for digital data transfer with a digital image.
With the introduction of error-correcting code and a novel digital signal processing technique suitable for nanomechanical systems, a nanoscale digital receiver with a vibrational nano-antenna composed of a carbon nanotube cantilever was experimentally shown to be capable of digital transfer.
In fact, the developed nanomechanical receiver successfully receive a digital image.
Furthermore, our fabrication method achieved a tiny gap around field emitter of vibrational nano-antenna, which enables the receiver to work with quite a low power consumption, on the order of 10\,nW.
Such an ultra-small, robust, and ultra-low-powered nanomechanical digital receiver opens up a new paradigm of digital data transfer at the nanoscale; collecting much more meaningful data via such device is expected to contribute to the forthcoming data-oriented age with IoT and AI-based systems.
\end{abstract}

%%%%%%%%%%%%%%%%%%%%%%%%%%%%%%%%%%%%%%%%%%%%%%%%%%%%%%%%%%%%%%%%%%%%%%%%%%%%%%%%%%%%%%%%%%%%%%%%%%%%%%%%%%%%%%%%%%%

The wireless transfer of digital data, such as text messages, pictures, and movies, enriches our modern life. 
The collection and analysis of scientific data with machine intelligence techniques have become essential techniques in natural and social science fields. 
Wireless communication terminals that have been scaled down to the nanoscale are expected to offer a more flexible platform for data collection\cite{Rizwan2018,Akyildiz2010}.
Such devices would enable promising future applications, including in the realm of healthcare with real-time in-body sensing, spatially distributed invisible camera, and surgery using nano-machines\cite{Nobunaga2018, Rizwan2018, Niu2019, Gardasevic2017}.
To develop such a nanoscale terminal, the front-end circuit of the terminal, including radio frequency\,(RF) antenna, should be downsized to the nanoscale\cite{Atakan2010}. 
Such components are often designed based on electromagnetic theory; the size of an RF antenna is designed on the order of the wavelength, which is the main issue during downsizing.

Nanomechanical systems offer a versatile platform for fundamental science and applications\cite{Roukes2001, Bunch2007, Sazonova2004, Jensen2008, Lee2010, Moser2013,Moser2014, Puller2013, Jensen2007, Hatanaka2017, Mahboob2008, Torrejon2017, Eom2011,Kosaka2014,  Chu2017, Hong2012, Connell2010, Singh2014, Tavernarakis2018}.
By exploiting the resonating vibration of a nanoscale cantilever, it has been reported that electromagnetic waves with frequency modulation\,(FM) can be received in nanoscale\cite{Jensen2007}.
The concept of receiving digitally modulated signals with nanomechanical systems has also been proposed\cite{tadokoro2018, tadokoro2018scirep}.
However, previously proposed receivers cannot obtain the entire energy of the incoming wave, which often causes error in the data transfer\cite{tadokoro2018}; to the best of our knowledge, no successful practical demonstration has yet been reported.

Here, we provide the first experimental demonstration of an ultra-small, robust, and ultra-low-powered digital receiver.
We achieved it by introducing concepts from recently developed physics of nanomechanics and original digital signal processing method.
Fortunately, sensed data is often expressed in a digital format, and thus, the use of error-correcting code should contribute to error reduction. 
Our original digital signal processing technique that is suitable for nanomechanical systems is developed, and it is shown to further enhance the error correction capability of the receiver.

\begin{figure}
	\centering
	\includegraphics[width=\columnwidth, keepaspectratio]{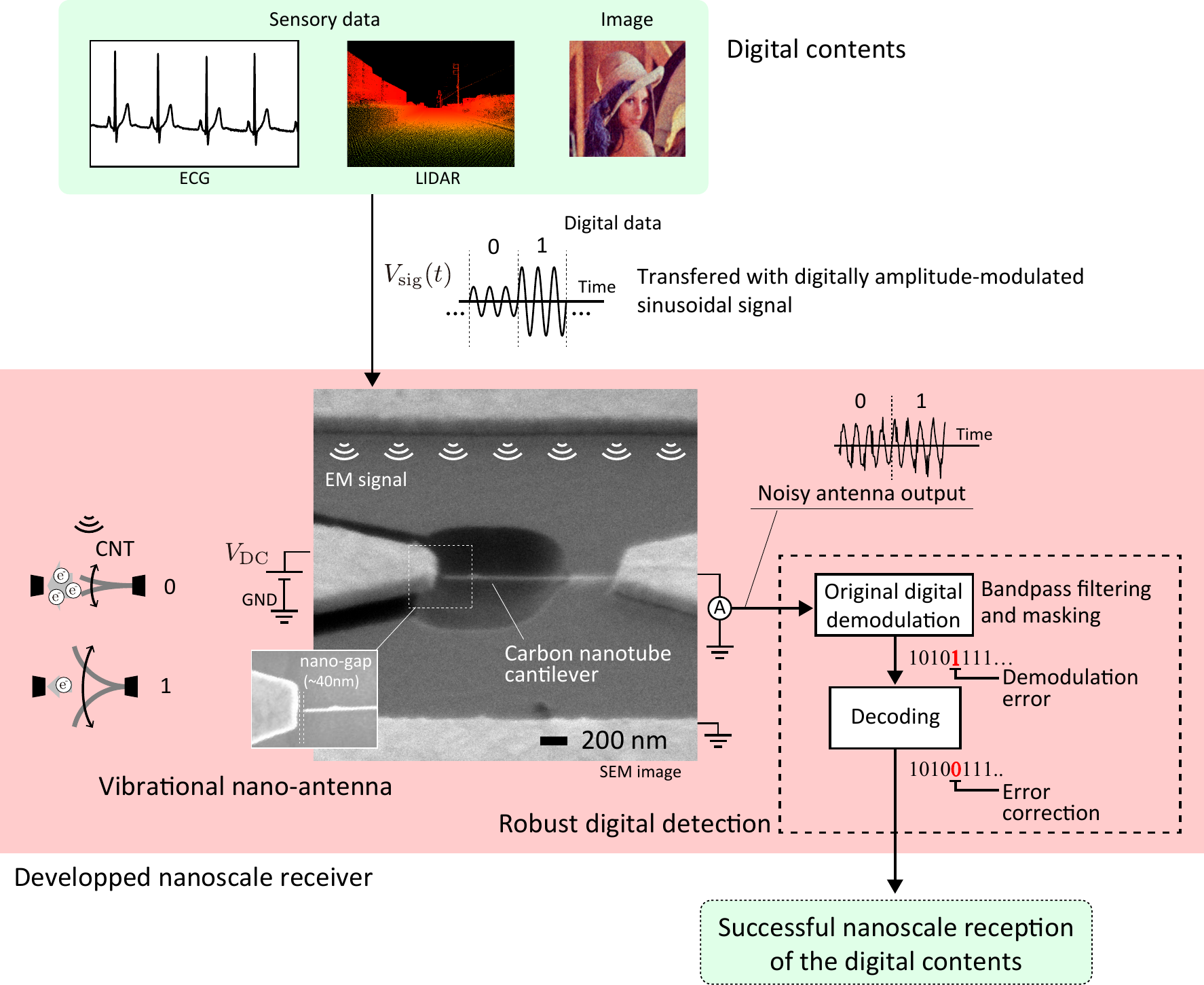} 
	\caption{
    Schematic of the developed digital receiver.
    This ultra-small nanomechanical digital receiver enables the robust transfer of digital data with ultra-low power consumption.
    The vibrational nano-antenna, which is composed of a CNT cantilever\,(see a scanning electron microscopy\,(SEM) image), significantly contributes to the miniaturisation of the RF part of the receiver.
    The resulting antenna output, however, is very noisy because of the nanoscale reception of the transmitted signal.
    A robust detection method, which involves original demodulation by bandpass filtering and masking, in addition to error correction, is used for signal discrimination.
    The introduced fabrication method achieves a tiny gap between field emitter\,(the tip of the cantilever) and counter-electrode, which enables the receiver to work with quite a low power consumption, on the order of 10\,nW.
    The first experimental demonstration of nanoscale digital transfer was successfully performed with the proposed system\,(Fig.\,\ref{fig:result}).
	}
	\label{fig:system}
\end{figure}

The developed nanomechanical digital receiver, which is presented in Fig.~\ref{fig:system}, was designed to be ultra-small, robust, and show ultra-low power consumption.
This receiver detects high-frequency signals, which can be used as carriers of digital contents, including digital images and sensory data.
In the developed receiver, the incoming electromagnetic\,(EM) signal excites vibrations in a nanomechanical system consisting of a carbon nanotube\,(CNT) cantilever\cite{Jensen2007, tadokoro2018, Atakan2010}.
Simply observing the vibrations gives information about the signal, including its amplitude and phase.
With the application of binary amplitude shift keying\,(BASK)\cite{proakis2001} as a digital modulation method, the amplitude level of the vibration itself can be used to represent binary digital information.
A sinusoidal signal with a small amplitude $A_0$ represents a binary value of 0 and induces a small vibration.
A large vibration comes from a signal with a large amplitude of $A_0\,{+}\,\Delta A$, which represents a binary value of 1\,(see Methods).
The term $\Delta A$ is a crucial variable that contributes to the discrimination of the transmitted data on the receiver side and is hereafter referred to as the ``signal gain''.
This vibrational method can be used to realise ultra-small antennas with sizes as small as 1.0\,$\mu$m.
Note that according to electromagnetic theory, the size of a traditional antenna that can receive a signal with a frequency of $f_\mathrm{c}\,{=}\,50$\,kHz is on the order of centimetres.

To reduce the detection error, a robust digital detection method was also introduced.
Because of the nanoscale reception of the signal, the vibration sometimes cannot follow the variation of the amplitude.
One typical situation is shown in Fig.\,\ref{fig:system}; despite the variation in the signal, the output current of the nano-antenna does not change, which makes it difficult to correctly estimate the transmitted data.
Although the demodulation method, which includes bandpass filtering and masking of the output current, reduces the amount of erroneously estimated data\,(see Supporting Information), it is unable to eliminate all detection error.
For more reliable detection, a robust encoding scheme of convolutional coding was applied with a coding rate of $R_\mathrm{C}\,{=}1/2$ and a constraint length of seven bits. 
Even with the demodulation error, in a decoding process with Viterbi decoding\cite{proakis2001}, the erroneously estimated bits are often corrected.
The details of the encoding, modulation, demodulation, and decoding are provided in the Methods section.
Note that detection process including decoding is often implemented in digital circuits.
Because such circuits are easily miniaturised with the current nanotechnology, employing our detection method does not affect the size of the receiver.

A concrete observation method is necessary to measure the amplitude of the vibration.
Similar to the frontier work by Jensen et al.\cite{Jensen2007}, the developed receiver uses electric current for the measurement.
When a voltage is applied, field emission occurs, and current flows along the CNT.
Unlike fabrication methods using chemical vapour deposition\cite{Moser2013, Moser2014, Jensen2007}, the present fabrication method based on semiconductor processes enables the inclusion of a tiny gap of 40\,nm between an electrode and the tip of the CNT cantilever\,(field emitter). 
The Fowler--Nordheim theory\cite{Fowler1928, Funayama2019} states that this tiny gap strongly enhances the current, meaning it becomes measurable even with our simple measurement system\,(see Methods).
In fact, in a vacuum of $8.0\,{\times}\,10^{-6}$\,Pa, applying a low voltage of $V_\mathrm{DC}\,{=}\,7.91$\,V allows the observation of a measurable current ${\gtrsim}\,1\,$nA\,(see Supporting Information).
These characteristics mean that the developed nanoscale receiver achieves ultra-low power consumption, on the order of 10\,nW for RF front-end.
Furthermore, because the gap width varies according to the amplitude of the vibration, the vibration amplitude can be estimated by simply observing the current.
This method can be easily implemented in electrical circuits even at the nanoscale, which improves the productivity of the developed nanomechanical digital receiver.

%%%%%%%%%%%%%%%%%%%%%%%%%%%%%%%%%%%%

\begin{figure}
	\centering
	\includegraphics[width=\columnwidth, keepaspectratio]{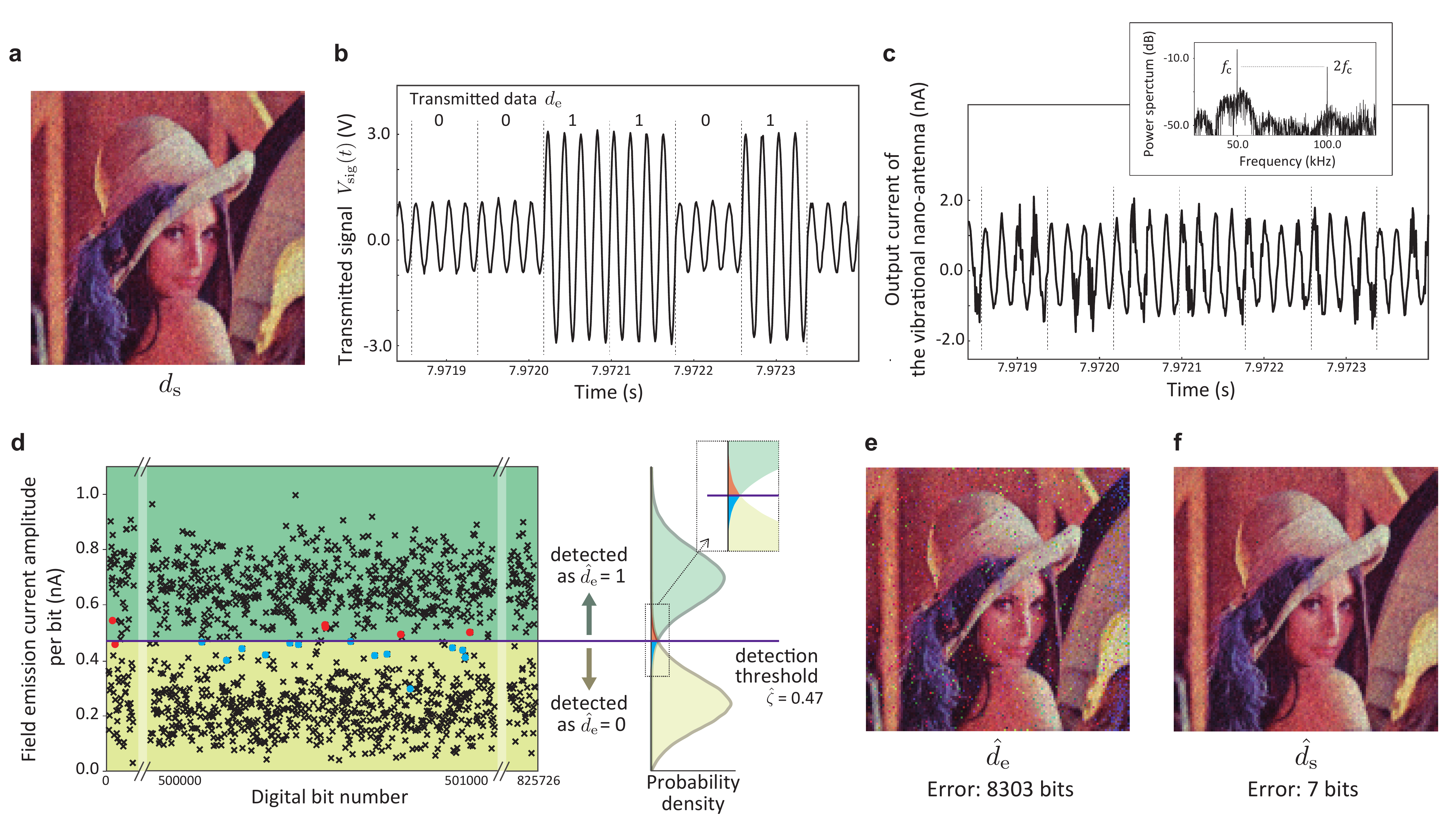} \\
	\caption{
    Experimental demonstration of the transfer of digital contents with a nanomechanical digital receiver.
    {\bf a}, Original digital image $d_\mathrm{s}$\,(full-color, 128\,$\times$\,128 pixels). {\bf b}, Part of the transmitted data $d_\mathrm{e}$ obtained after the encoding and digital modulation, and the corresponding signal $V_\mathrm{sig}(t)$. 
    {\bf c}, Output current of the vibrational nano-antenna\,(field emission current). The inset shows the frequency characteristics; the desired component excited by the vibration is the frequency $2f_\mathrm{c}$, not $f_\mathrm{c}$. 
    {\bf d}, Amplitude $A_\mathrm{DC}$ of the current samples in a part of the transmitted data. These samples were obtained after our original digital detection, namely, bandpass filtering and masking. They contain erroneously estimated bits, which have estimated values that are not equal to the transmitted values\,(coloured points). The probabilities of the erroneous detection are the area with blue and magenta. 
    {\bf e}, Estimated digital image obtained by the receiver; in $N_\mathrm{e}\,{=}\,$393,216 transmitted bits, detection error occurred in 8,303 bits, yielding an imperfect image. {\bf f}, Corrected image obtained after the decoding process. Almost all of the errors were corrected; only 7 bits with the detection error remain in the corrected image. 
 %\textcolor{red}{Note: the images (e) and (f) were obtained in different measurement. }
	}
	\label{fig:result}
\end{figure}

It was experimentally demonstrated that the developed nanoscale receiver could successfully detect a transmitted digital image\,(Fig.\,\ref{fig:result}).
The original image is RGB full-colour, 128\,$\times$\,128\,pixels, and composed of $N_\mathrm{e}\,{=}\,$393,216\,bits\,(Fig.\,\ref{fig:result}a).
The image data were encoded, and modulated to obtain the transmitted signal $V_\mathrm{sig}(t)$\,(Fig.\,\ref{fig:result}b, see Methods).
In the experiment, the amplitude parameters were set to $A_0\,{=}\,$1.0\,V, $\Delta A\,{=}\,$2.0\,V.
The digital image obtained by the nanomechanical digital receiver is shown in Fig.\,\ref{fig:result}f.
Although the output current of the vibrational nano-antenna cannot describe the amplitude level of the transmitted signal\,(Fig.\,\ref{fig:result}c), the original image was successfully received\,(Fig.\,\ref{fig:result}f).
This robustness arises from the proposed digital detection method: bandpass filtering and masking, combined with error correction.
In this detection, bandpass filtering with a passband centred on $f\,{=}\,2f_\mathrm{c}$ is first applied to the output current.
The results reveal that, as shown in the inset of Fig.\,\ref{fig:result}c, the current has two peaks in the power spectrum at $f\,{=}\,f_\mathrm{c}$ and $2f_\mathrm{c}$.
The current resulting from the vibration has a frequency of $f\,{=}\,2f_\mathrm{c}$, see Ref\cite{tadokoro2018}.
The other component is a capacitive current arising because of the close placement of the signal electrodes to the vibrational nano-antenna\,(Extended Data Fig.\,1).

A unique procedure in the proposed detection method is the masking of the output current.
Figure\,\ref{fig:result}b indicates that four periods of the sinusoidal signal describe a single data bit.
The observation results indicate that nearly the first half of the signal\,(i.e., the first two periods) are affected by the previous data bit\,(see Supporting Information).
This erroneous effect comes from a transient behaviour in response to the data stream switching from 0 to 1 or vice versa.
In these cases, the amplitude of the signal is digitally changed; however, because of the inertial effects, the tip of the CNT cantilever cannot immediately respond to the switching.
As a result, the estimation of the transmitted data is incorrect, i.e., $d_\mathrm{e}\,{\neq}\,\hat{d}_\mathrm{e}$.
To avoid the erroneous effects of this peculiar behaviour of the vibrational nano-antenna, only the last two periods were used in the detection.
The details of the proposed digital detection method are provided in the Methods.

Even with this digital detection method, detection error still remains.
The amplitude of the current after filtering and masking, $A_\mathrm{DC}$, is shown in Fig.\,\ref{fig:result}d to depict the detection error before error correction.
Here, a detection threshold was introduced to determine the binary value of the transmitted data; if the current is below the threshold, the transmitted bit is estimated to be 0.
The plot shows two distinct clusters of the points, with one having a large amplitude corresponding to bits with $d_\mathrm{e}\,{=}\,1$ and the other corresponding to $d_\mathrm{e}\,{=}\,0$.
Although most of the points correctly describe the amplitude of the transmitted signal\,(marked with black points), a few cases result in error\,(marked with coloured points).
This observation is visually understandable from the probability density for the currents shown on the right in Fig.\,\ref{fig:result}d\,(see Methods).
The green and yellow area represent the probability of detection being correct, i.e., the correct detection rate $P[\hat{d}_\mathrm{e}=d_\mathrm{e}]$.
These two areas comprise most of the probability density, which indicates that most of the transmitted data are correctly estimated in the developed nanomechanical digital receiver.
This success arises from the setting of the threshold value.
To minimise erroneous detection, the probabilities for these two regions are calculated, and the threshold is set to make these two probabilities equal\,(see Methods).
Despite the use of this optimal threshold, some level of error is unavoidable, as shown in Fig.\,\ref{fig:result}d.
The estimated image before error correction\,(Fig.\,\ref{fig:result}e) shows that some pixels in the received image collapsed.
A comparison of the images before and after error correction\,(Fig.\,\ref{fig:result}e and f, respectively) demonstrates that the error correction strongly contributes to correct data transfer.

%%%%%%%%%%%%%%%%%%%%%%%%%%%%%%%%%%%%%%%%%%%%%%%%%%%%%%%%%%%%%

\begin{figure}
	\centering
	\includegraphics[width=\columnwidth, keepaspectratio]{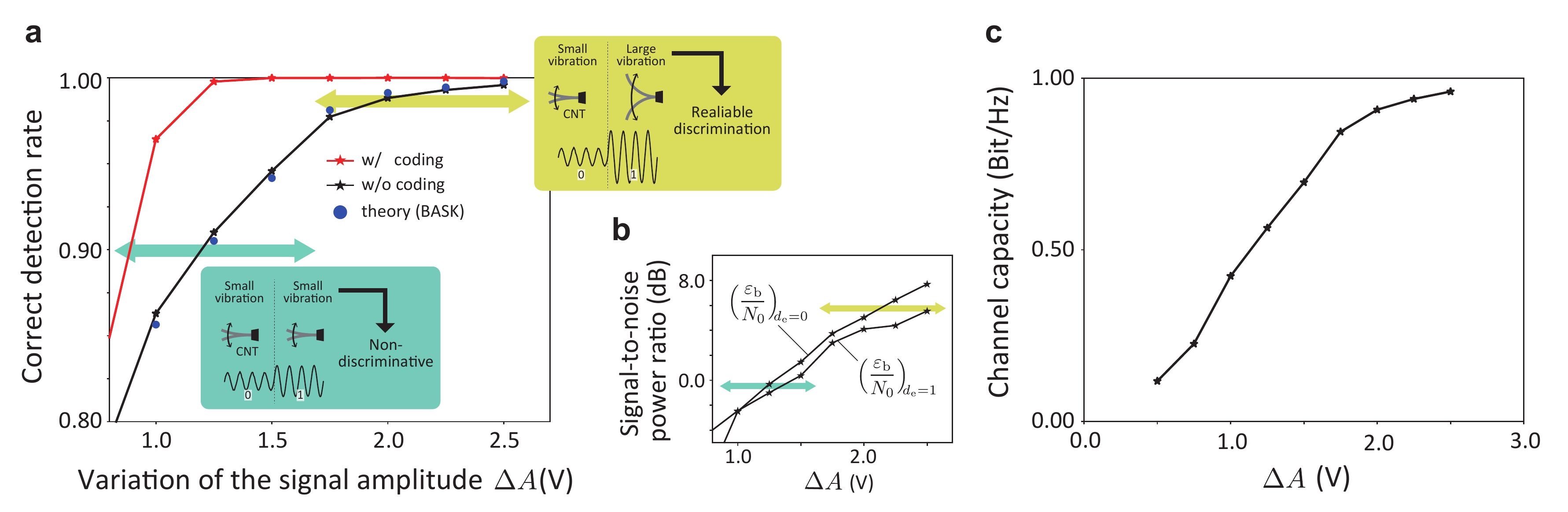}
	\caption{
	Communication performance of the developed nanomechanical digital receiver: 
	{\bf a}, Correct detection rate. Decreasing $\Delta A$ results in a lower rate of the correct detection. Even with the demodulation error, the rate can be kept to be high in the region where $\Delta A\,{\gtrsim}\,1.25$\,V, because of the error correction method. The theoretical value is also shown, as calculated from the S/N per bit, which is plotted in {\bf b}\,(see Methods). The experimental results obtained without coding agree well with the theoretical results.
	{\bf b}, S/N per bit, $\varepsilon_\mathrm{b}/N_0$.  
	Because the variance of the probability density depends on the value of the transmitted data $d_\mathrm{e}$\,(see Methods), the ratio for $d_\mathrm{e}=0$ and $1$ are slightly different.
    {\bf c}, Channel capacity. With sufficiently large $\Delta A$, the receiver achieved a capacity of 1.0\,bit/Hz, which is the theoretical upper-bound in BASK modulation\cite{proakis2001}.
	}	
	\label{fig:BER}
\end{figure}

It was found that increasing the signal gain $\Delta A$ contributes to the reduction of erroneous detection.
Figure\,\ref{fig:BER}a shows the dependence of the correct detection rate on the gain.
To measure the performance of the receiver in this regard, the reception of the transmitted signal was observed for randomly generated input digital data, $d_\mathrm{s}{\in}\{0,1\}$.
When the gain was sufficiently large\,($\Delta A\,{\gtrsim}\,2.0$), the receiver was able to correctly discriminate the transmitted data in the presence of noise.
Such a gain realises a significant difference in the vibrational amplitudes corresponding to the two states, which contributes to accurate discrimination.
In the case of a small gain, however, the receiver fails to recover many transmitted data, even with the error correction.

The signal-to-noise power ratio\,(S/N), a key performance indicator in terms of signal processing or communication systems, was also evaluated in the nanoscale receiver.
From the measurement results presented in Fig.\,\ref{fig:result}d, the ratio per bit was calculated\,(see Methods).
The two curves for the S/N in Fig.\,\ref{fig:BER}b correspond to the transmitted data with $d_\mathrm{e}{=}0$ and $d_\mathrm{e}{=}1$.
It is well-known that for reliable communication, the error rate should be smaller than the order of $10^{-3}$.
Our receiver achieves this condition when the S/N is larger than approximately 6\,dB. 
Employing the error correction method enhances the ability of the receiver to secure a high S/N.
Based on this performance, the theoretical value of the correct detection rate can be calculated\,(see Methods).
The blue circles in Fig.\,\ref{fig:BER}a show these results; the experimental performance of the developed receiver without the error correction agrees well with the theoretical one.

The limit on wireless transfer of digital data with the receiver was also investigated.
In the field of information theory, a performance measure called channel capacity, which characterises the rate at which digital information can be reliably transmitted over a wireless channel, has been introduced\cite{proakis2001}.
A large channel capacity indicates that the capability of high-speed transfer and that the use of the bandwidth is efficient.
The channel capacity of the developed receiver was calculated from the measured probability density\,(see Methods).
As shown in Fig.\,\ref{fig:BER}c, the developed nanoscale receiver achieves a capacity of 1.0\,bit/Hz, which is the theoretical upper bound in BASK modulation\cite{proakis2001}.
This high-efficiency, reliable reception indicates that even a nanoscale receiver has the capability of high-speed transfer; for example, it could achieve a rate of 10\,Mbps for a digitally modulated signal with a bandwidth, 10\,MHz, which is a common situation in WiFi systems.

%%%%%%%%%%%%%%%%%%%%%%%%%%%%%%%%%%%%%%%%%%%%%%%%%%%%%%%%%%%%%

The achievements in this work represent a major milestone in the development of nanoscale digital wireless terminals.
This Letter reports the first practical demonstration that nanomechanical systems are capable of detecting digitally modulated signals.
With the use of a vibrational nano-antenna, the proposed robust digital signal processing method enables reliable nanoscale data transfer.
The developed fabrication method contributes to strongly reducing the power consumption.
The present experiments demonstrate that a digital image carried by a modulated signal can be successfully detected with the developed nanomechanical system.
Such an ultra-small, robust, and ultra-low-powered nanomechanical digital receiver opens up a new paradigm of digital data transfer at the nanoscale, which will contribute to the forthcoming age of {\it true} IoT world: everything are connected.

%\# 1391 words after the summary.

%Formatting notes:
%\vspace{-\baselineskip}
%\begin{itemize}
%    \item The rest of the text(after the summary) is typically about 1,500 words long (not including Methods, summary paragraph or other sections).
%    \item As a guideline they allow up to 30 references.
%    \item Spelling must be British English (Oxford English Dictionary).
%\end{itemize}

\newpage

% script:
%  20190707.draw_modulated_signal.Fig1_Vsig.ipynb
%  20190708.Nature_Nanoreceiver.Fig1_Fig2d.ipynb
% Picture:
%  SEM Original tilt.tif
%  SEM Original.jpg
%  CNT.png

\newpage

% script:
%  20190718.Nature_Nanoreceiver_tad.Fig2b_Fig2c.ipynb
%  20191113.Nature_Nanoreceiver_tad.Fig2c_spectrum.ipynb
%  20190708.Nature_Nanoreceiver.Fig1_Fig2d.ipynb
%  
% Picture:
%  a: Lenna_s128_06.png
%  e: Fig2e_lena_wo.eps
%  f: Fig2f_decoded_lena_3V_1V.png

\newpage

% script:
%  20190830.Nature_BER.Fig3a.ipynb
%  230191118.Nature_BER_theory_tad.Fig3b.ipynb
%  20190828.Nature_channelCapasity.Fig3c.ipynb
%  

\bibliographystyle{naturemag}

%% Here is the endmatter stuff: Supplementary Info, etc.
%% Use \item's to separate, default label is "Acknowledgements"

\begin{addendum}
 \item Part of this work was supported by Nagoya University microstructural characterization platform as a ``Nanotechnology Platform'' program of the Ministry of Education, Culture, Sports, Science and Technology\,(MEXT), Japan.
 \item[Competing Interests] The authors declare that they have no
competing financial interests.
 \item[Correspondence] Correspondence and requests for materials should be addressed to Y.T.\\(email:\,tadokoro@mosk.tytlabs.co.jp).
 \item[Author Contributions] K.\,F. designed and fabricated the nanoscale receiver, with the assistance of K.\,S, J.\,H. and Y.\,O.
H.\,T. and Y.\,T. helped to build the signal processing method with K.\,F.
All authors co-wrote the paper.
Y.\,T. and Y.\,O. supervised this project.
 \end{addendum}

%%%%%%%%%%%%%%%%%%%%%%%%%%%%%%%%%%%%%%%%%%%%%%%%%%%%%%%%%%%%5
\newpage

\begin{methods}
%Put methods in here.  If you are going to subsection it, use \verb|\subsection| commands. 
%Methods section should be less than 800 words and if it is less than 200 words, it can be incorporated into the main text.

%\subsection{Mechanism of the vibration.}

%An electrostatic force vibrates the tip of the CNT cantilever\cite{Jensen2007, tadokoro2018}.
%Applying a voltage $V_\mathrm{DC}$ induces an electrical charge $Q_\mathrm{DC}$ at the tip of the CNT\cite{Tanaka2015}.
%The signal $V_\mathrm{sig}(t)$, which is applied at Signal electrode \#1 in Extended Data Fig.\,\ref{fig:device}, induces an electromagnetic\,(EM) field, $E_\mathrm{sig}(t)$, around the tip.
%In this situation, according to Coulomb's law, the electrostatic force $F\,{=}\,Q_\mathrm{DC}E_\mathrm{sig}(t)$ is excited at the tip.
%Since the driving force describes the digital signal itself, observing the vibration gives us the information of the signal, such as amplitude and phase.

\subsection{Preparing the transmitted digitally modulated signal.}

Digital data $d_\mathrm{s}\,{\in}\,\{0,1\}$ with $N_\mathrm{s}$ bits were transmitted to the nanomechanical digital receiver via an electromagnetic digitally modulated signal $V_\mathrm{sig}(t)$.
The first step of preparing the signal is the encoding of the data for error correction.
Here, a standard error collecting code was employed: a convolutional code with a constraint length of seven bits and a coding rate of $R_\mathrm{C}\,{=}\,1/2$\,(see Ref\cite{proakis2001}).
In the experiment, to obtain the encoded data, the two functions `{\tt poly2trellis}' and `{\tt convenc}', which are provided in MATLAB\textregistered 2012b, were used. 
The encoded data $d_\mathrm{e}\,{\in}\,\{0,1\}$ has a size of $N_\mathrm{e}\,{=}\,N_\mathrm{s}/R_\mathrm{C}$.
These data are then mapped to the corresponding time-continuous high-frequency sinusoidal signal $V_\mathrm{sig}(t)$.
According to BASK modulation, the data $d_\mathrm{e}\,{=}\,0$ are mapped to the carrier signal with a small amplitude $A_0\,{>}\,0$, and the data of $d_\mathrm{e}\,{=}\,1$ are mapped to that with a large amplitude $A_0\,{+}\,\Delta A$.
Each encoded bit is expressed by a signal with time-duration, $T_\mathrm{b}$.
To reflect this point, a single pulsed signal $g_\tau(t)\,{=}\,1$ for $0\,{\leq}\,t\,{\leq}\,\tau$ and in the other time-interval, $g_\tau(t)\,{=}\,0$, is introduced.
The digitally modulated signal in this case is theoretically described by
\begin{align}
V_\mathrm{sig}(t) = \sum_{j_\mathrm{e}=0}^{N_\mathrm{e}-1} ( A_0 + \Delta A d_\mathrm{e}[j_\mathrm{e}])g_{T_\mathrm{b}}(t-j_\mathrm{e}T_\mathrm{b}) \cos 2\pi f_\mathrm{c}t .
\label{eq:signal}
\end{align}
In the experiment, $f_\mathrm{c}$ was set to 50\,kHz and four periods of the sinusoidal signal were used to describe one bit, giving $T_\mathrm{b}\,{=}4/f_\mathrm{c}$\,(see Fig.\,\ref{fig:result}b).

\subsection{Device fabrication and structural features.}

The developed device was fabricated with a standard semiconductor process.
Extended Data Fig.\,1 shows the structure of the developed device.
The emitter\,(cathode) consists of a singly clamped string constructed by applying multiwall CNTs to a silicon wafer. 
This was accomplished by dropping a dispersion of CNTs in isopropyl alcohol onto a silicon wafer with a 3\,$\mu$m-thick silicon dioxide layer grown by plasma-enhanced chemical vapour deposition. 
The solvent was subsequently evaporated using a heating device. 
The straight CNTs used in this work were originally synthesised by an arc discharge process and generally had a consistent structure in terms of length, radius and tip shape. 
A Au electrode was attached to one end of the CNT string using electron beam lithography, metal deposition, and lift-off processes, and a second Au electrode was deposited such that there was a tiny gap $h_0$ of approximately 40\,nm from the other end of the string.
The resulting CNT cantilever had a length and radius of $L\,{=}\,900$\,nm and $\rho\,{=}\,19.0$\,nm, respectively. 
At the same time, two signal electrodes were placed using electron beam lithography.
The distance between them was $D_\mathrm{s}\,{=}\,3.0\,\mu$m, and they were separated from the anode and cathode with a distance of at least $D_\mathrm{c}\,{=}\,1.0\,\mu$m.
The CNT string was suspended by etching the underlying silicon oxide to a depth of 500\,nm using hydrofluoric acid and was released in a critical point dryer. 
The width of both the anode and cathode in the test specimens was $w\,{=}\,$300\,nm, and the height of each electrode was approximately 110\,nm. 
Note that the above lengths, distances and radii were measured via SEM images.

\subsection{Measurement set-up for observing the nanomechanical digital receiver.}

The transmitted digital information $d_\mathrm{s}$ is estimated by detecting the signal $V_\mathrm{sig}(t)$ in the nanomechanical digital receiver.
Extended Data Fig.\,2 shows the experimental setup for observing the behaviour of the receiver during signal detection.
The fabricated sample was placed in a vacuum chamber with a vacuum level of $8.0\,{\times}\,10^{-6}$\,Pa at a temperature of 298\,K. 
Between the anode and cathode, a constant voltage of $V_\mathrm{DC}\,{=}\,7.91$\,V was applied to excite an electronic charge around the tip of the CNT cantilever.
This voltage was supplied by a regulated direct current power supply\,(PMX110-0.6A, Kikusui Electronics Corporation).
Under these conditions, the signal $V_\mathrm{sig}(t)$ was applied to signal electrode\,\#1.
Because the opposite electrode, signal electrode\,\#2, was grounded, the signal induced an electric field between the electrodes, including around the cantilever.
In the experiment, according to Eq.\,\eqref{eq:signal}, the signal was generated by a arbitrary waveform generators\,(Mi2.6021, SPECTRUM) with control software\,(SBench6-Pro\,6.1).
All of the cables were electrically isolated from the vacuum chamber to avoid any crosstalk between the signal line\,(signal electrodes\,\#1 and \#2) and the current measurement line\,(anode and cathode).

The electric current in nanomechanical systems, such as the field emission current, is weak (generally on the order of 1\,nA).
This provides a motivation for the use of a lock-in-amplifier in the measurement\cite{Jensen2008, Sazonova2004, Moser2014}.
Unfortunately, the current output from the vibrational nano-antenna is a time-varying signal with a rate of $T_\mathrm{b}$.
This indicates that such a measurement device is not valid.
Thus, an alternative method was introduced: high-frequency digital sampling of the current and post-processing with the stored sampled data.
First, the current was converted into a voltage $V_\mathrm{R}(t)\,{=}\,RI(t)$ with a resistance of $R\,{=}\,100\,\mathrm{k}\Omega$.
The voltage was amplified by a low-noise preamplifier\,(LI-75A, NF Corporation) with a gain of $G\,{=}\,40$\,dB.
Here, the input impedance of the amplifier is sufficiently large compared to $R$.
Then, the amplified voltage was measured by a digital oscilloscope\,(Digital phosphor oscilloscope, DPO7104C, Tektronix).
This voltage was sampled at a rate of $f_\mathrm{s}\,{=}\,1.0$\,MHz.
The sampled data $V_\mathrm{rcv}$ were saved in a memory device.
The post-processing for including bandpass filtering and error correction was performed on a desktop personal computer\,(PC).

\subsection{Robust digital detection.}

From the measured data $V_\mathrm{rcv}$, the developed receiver estimates the transmitted data $d_\mathrm{s}$ according to the proposed detection method.
The measured current is first set to $I_\mathrm{rcv}\,{=}\,V_\mathrm{rcv}/GR$.
Next, the current is masked to avoid transient effects.
Our observation revealed that almost the first half of the segment of the signal representing a single bit, i.e., the first two periods, are affected by the previous bit.
When the amplitude of the signal changes, this effect leads to erroneous estimation\,(see Supporting Information).
To counteract this effect, the first half of the current is deleted for every bit, and the remaining part, denoted $\tilde{I}_\mathrm{rcv}$, is used in the detection.

Next, bandpass filtering is applied to $\tilde{I}_\mathrm{rcv}$ to extract the $2f_\mathrm{c}$ frequency component.
After $\tilde{I}_\mathrm{rcv}$ is multiplied by the reference signal $e^{i(2\pi\cdot2f_\mathrm{c}t+\theta_\mathrm{r})}$, a three-step filtering procedure is applied: 1) taking the discrete Fourier transform~(DFT) of $\tilde{I}_\mathrm{rcv}e^{i(2\pi\cdot2f_\mathrm{c}t+\theta_\mathrm{r})}$ to calculate the Fourier coefficients for each frequency; 2) replacing all coefficients with zero, except that corresponding to zero frequency, to eliminate high-frequency components; and 3) taking the inverse DFT\,(IDFT) with the replaced coefficients to reproduce the filtered current samples.
The resulting current is given by $\hat{I}_\mathrm{rcv}\,{=}\,A_\mathrm{DC}e^{i(\theta_\mathrm{s}-\theta_\mathrm{r})}$, where $\theta_\mathrm{s}$ and $\theta_\mathrm{r}$ are the phases of the reference signal and the $2f_\mathrm{c}$ component of the current, respectively.
In the estimation of the transmitted data, the absolute value of the current, $|\hat{I}_\mathrm{rcv}|$, is used.
The current is also plotted in Fig.\,\ref{fig:result}d.

As shown in Fig.\,\ref{fig:result}d, the transmitted data $d_\mathrm{e}$ are estimated by comparing $A_\mathrm{DC}$ with a threshold, $\zeta$: the estimated data $\hat{d}_\mathrm{e}$ are obtained as $\hat{d}_\mathrm{e}\,{=}\,1$ when $A_\mathrm{DC}\,{>}\,\zeta$ and $\hat{d}_\mathrm{e}\,{=}\,0$ otherwise.
To maximise the probability of correct detection, the optimal threshold $\hat{\zeta}$ is selected, described by $\hat{\zeta} = \mathrm{argmax}_{\zeta}\big[ P[ \hat{d}_\mathrm{e}\,{=}\,1|d_\mathrm{e}\,{=}\,1] + P[\hat{d}_\mathrm{e}\,{=}\,0|d_\mathrm{e}\,{=}\,0]\big]$.
%The probability $P[ A_\mathrm{DC}\,{>}\,\zeta|d_\mathrm{e}\,{=}\,0]$ is the area which is coloured with magenta in Fig.\,\ref{fig:result}d.
%The area which is coloured with blue indicates  $P[ A_\mathrm{DC}\,{<}\,\zeta|d_\mathrm{e}\,{=}\,1]$.
With the experimental data considered here, the optimum threshold calculated to be $\hat{\zeta}\,{=}\,0.47$, which is located at the crossing point of the two probability densities as shown in Fig.\,\ref{fig:result}d.
%This threshold achieves $P[ A_\mathrm{DC}\,{>}\,\zeta|d_\mathrm{e}\,{=}\,0]\,{=}\,P[ A_\mathrm{DC}\,{<}\,\zeta|d_\mathrm{e}\,{=}\,1]\,{=}\,xx$.

Even with the optimal threshold, some bits are still erroneously detected.
The error is then further reduced by introducing Viterbi decoding method\cite{proakis2001}.
The decoder, which corresponds to the convolutional encoding used to prepare the transmitted signal, is provided by MATLAB\textregistered 2012b.
The function `{\tt vitdec}' with the input sequence of $\hat{d}_\mathrm{e}$ is applied to obtain the final estimated data $\hat{d}_\mathrm{s}$, which is used to reproduce the final received image shown in Fig.\,\ref{fig:result}f.

\subsection{Drawing the probability density and estimating noise variance and signal-to-noise power ratio.}

The probability density is obtained by a powerful statistical tool called kernel density estimation\,(KDE)\cite{Silverman1986, Funayama2019}, which is a method of estimating the probability density function from a finite selection of observation values.
The probability density of the current $x$ for the transmitted data with $d_\mathrm{e}\,{=}\,1$, which corresponds to the green area in Fig.\,\ref{fig:result}d, is obtained from the experimental data of $A_{\mathrm{DC},d_\mathrm{e}{=}1}$ as $P[ x | d_\mathrm{e}\,{=}\,1 ] =\frac{1}{N_\mathrm{e}} \sum K \big(x - A_{\mathrm{DC},d_\mathrm{e}{=}1} \big)$, where $K(\cdot)$ is a Gaussian kernel function with zero mean and unit variance.
The operation $\sum(\cdot)$ in this equation indicates the sum taken over all $N_\mathrm{e}$ samples on $A_{\mathrm{DC},d_\mathrm{e}{=}1}$.
The probability density for the transmitted data with $d_\mathrm{e}\,{=}\,0$, corresponding to the yellow area in Fig.\,\ref{fig:result}d, is obtained similarly. 

We have previously reported that the probability density for the current follows a Gaussian distribution\cite{Funayama2019}.
With the experimental data, the two essential parameters that characterise the distribution, the mean and variance, were calculated as $\mu_{d_\mathrm{e}{=}1}\,{=}\,7.01{\times}10^{-10}$\,A and $\sigma_{d_\mathrm{e}{=}1}^2\,{=}\,1.11{\times}10^{-20}\,\mathrm{A}^2$ for $d_\mathrm{e}\,{=}\,1$ and $\mu_{d_\mathrm{e}{=}0}\,{=}\,2.52{\times}10^{-10}$\,A and $\sigma_{d_\mathrm{e}{=}0}^2\,{=}\,9.16{\times}10^{-21}\,\mathrm{A}^2$ for $d_\mathrm{e}\,{=}\,0$.
The result of $\sigma_{d_\mathrm{e}{=}1}^2\,{\neq}\,\sigma_{d_\mathrm{e}{=}0}^2$ indicates that the two probability densities have a different variance.

The properties of the distributions give us an important information on the communication performance of the receiver.
One is the S/N, which is widely used to evaluate the signal gain in engineering fields.
The distance from the threshold to the mean, which is given by $\varepsilon_{\mathrm{b}, d_\mathrm{e}{=}\{0,1\}}\,{=}\,|\mu_{d_\mathrm{e}{=}\{0,1\}}-\zeta|^2$, is equal to the signal power received at the nanomechanical digital receiver\cite{proakis2001, tadokoro2018}.
The variance is the two-sided power spectrum of the noise\cite{proakis2001}, as described by $N_{\mathrm{0}, d_\mathrm{e}{=}\{0,1\}}/2\,{=}\,\sigma_{d_\mathrm{e}\,{=}\,\{0,1\}}^2$.
These observations give the S/N, $(\varepsilon_\mathrm{b}/N_0)_{d_\mathrm{e}{=}\{0,1\}}\,{=}\,\varepsilon_{\mathrm{b}, d_\mathrm{e}{=}\{0,1\}}/N_{\mathrm{0}, d_\mathrm{e}{=}\{0,1\}}$.
The ratio for each bit depends on the value of the transmitted data $d_\mathrm{e}$ because the variances $\sigma_{d_\mathrm{e}{=}0}^2$ and $\sigma_{d_\mathrm{e}{=}1}^2$ are different for the two possible values.
The ratio is plotted in Fig.\,\ref{fig:BER}b.

The theoretical error rate can be calculated from the above S/N.
Traditional communication theory shows that the error rate in BASK with different variances is $Q(\sqrt{2(\varepsilon_\mathrm{b}/N_0)_{d_\mathrm{e}{=}0}})/2+Q(\sqrt{2(\varepsilon_\mathrm{b}/N_0)_{d_\mathrm{e}{=}1}})/2$ where $Q(x)\,{=}\,(1/\sqrt{2\pi})\int_x^\infty e^{-u^2/2}\mathrm{d}u$ is the {\it  Q-function}\cite{proakis2001}.
This theoretical value is also plotted in Fig.\,\ref{fig:BER}a.

\subsection{Calculation of channel capacity.}

The observations shown in Fig.\,\ref{fig:BER} demonstrate that the error rate depends on whether the transmitted bit takes a value of $d_\mathrm{e}\,{=}\,0$ or  $d_\mathrm{e}\,{=}\,1$ because the S/N is different in these two cases.
This behaviour means that the data transfer in the developed nanomechanical receiver is theoretically characterized by an asymmetric binary channel\cite{proakis2001, Chapeau1997, Tanaka2019}.
In this case, the channel capacity is derived by the maximizing the mutual information $I(d_\mathrm{e};\hat{d}_\mathrm{e})=H(\hat{d}_\mathrm{e})-H(\hat{d}_\mathrm{e}|d_\mathrm{e})$ with respect to the {\it a priori} probability $\alpha\,{\equiv}\,P[d_\mathrm{e}]$.
Here, $H(X)$ is the entropy of the random variable $X$, see Ref\cite{proakis2001}.
The maximum value of $I(d_\mathrm{e};\hat{d}_\mathrm{e})$ is obtained when $\mathrm{d}I/\mathrm{d}\alpha\,{=}\,0$, which gives $\alpha\,{=}\,\{1-p_1(1+z)\}/\{(1-p_0-p_1)(1+z)\}$ where $z\,{=}\,2^{\{h(p_0)-h(p_1)\}/(1-p_0-p_1)}$ and $h(p)$ is the binary entropy function.
Here, we have two probabilities, $p_0\,{=}\,P[\hat{d}_\mathrm{e}{=}1|d_\mathrm{e}{=}0]$ and $p_1\,{=}\,P[\hat{d}_\mathrm{e}{=}0|d_\mathrm{e}{=}1]$.
Substituting the obtained value of $\alpha$ into the equation for the mutual information gives the channel capacity: 
\begin{align}
c = \log_2(1+z) - \frac{1-p_1}{1-p_0-p_1}h(p_0) + \frac{p_0}{1-p_0-p_1}h(p_1).
\end{align}
The plots in Fig.\,\ref{fig:BER}c were obtained from the above expression.

\end{methods}

\newpage

\begin{figure}
	\centering
	\includegraphics[width=0.65\columnwidth, keepaspectratio]{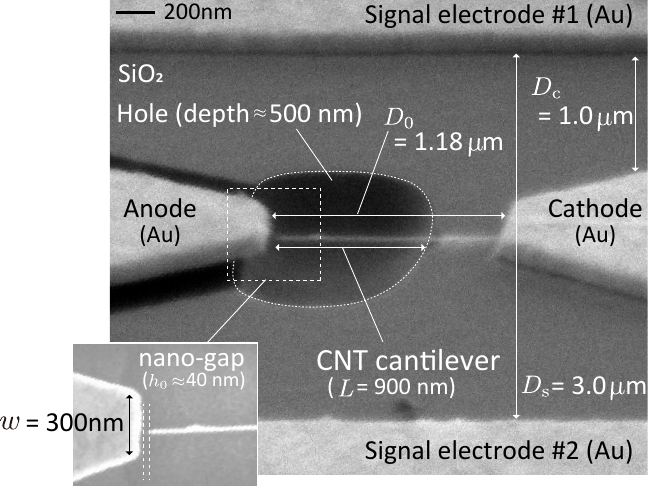} % this command will be ignored
%	\caption{ {\bf Extended Data Fig.\,1:} Structure of the fabricated sample device.}
    \\
    \begin{center}
        {\bf Extended Data Fig.\,1:} Structure of the fabricated sample device.
    \end{center}
	\label{fig:device}
\end{figure}

\newpage

\begin{figure}
	\centering
	\includegraphics[width=1.0\columnwidth, keepaspectratio]{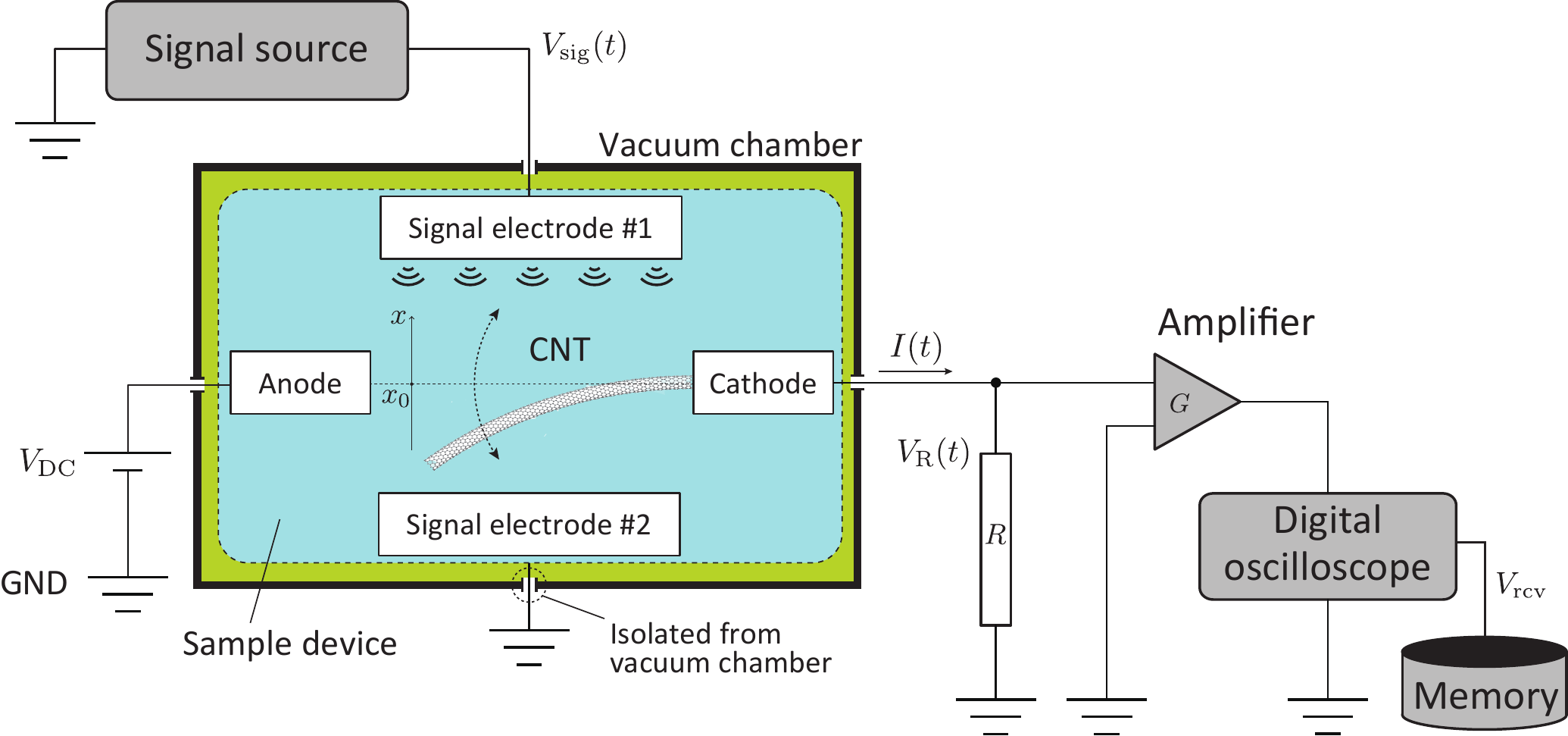} % this command will be ignored
%	\caption{ {\bf Extended Data Fig.\,2:} Measurement set-up for observing the behavior of the nanomechanical digital receiver.}
	    \\
    {\bf Extended Data Fig.\,2:} Measurement set up for observing the behaviour of the nanomechanical digital receiver.
	\label{fig:overall_system}
\end{figure}

%\bibliographystyle{naturemag}
%\bibliography{ref_method}

%% Put the bibliography here, most people will use BiBTeX in
%% which case the environment below should be replaced with
%% the \bibliography{} command.

%\renewenvironment{figure}{\let\caption\NAT@figcaption}{}

%%
%% TABLES
%%
%% If there are any tables, put them here.
%%

%\begin{table}
%\centering
%\caption{This is a table with scientific results.}
%\medskip
%\begin{tabular}{ccccc}
%\hline
%1 & 2 & 3 & 4 & 5\\
%\hline
%aaa & bbb & ccc & ddd & eee\\
%aaaa & bbbb & cccc & dddd & eeee\\
%aaaaa & bbbbb & ccccc & ddddd & eeeee\\
%aaaaaa & bbbbbb & cccccc & dddddd & eeeeee\\
%1.000 & 2.000 & 3.000 & 4.000 & 5.000\\
%\hline
%\end{tabular}
%\end{table}

\end{document}